\documentclass[preprint,12pt]{elsarticle}

\usepackage{amssymb}
\usepackage{amsmath}

\usepackage{lineno}

\usepackage{orcidlink}
\usepackage{svg}
\usepackage{tabularx}
\usepackage{rotating} 
\usepackage{siunitx}  

\DeclareSIUnit\byte{B}
\DeclareSIUnit{\clight}{\text{\ensuremath{c}}}
\DeclareSIPrefix{\kibi}{Ki}{10}
\DeclareSIPrefix{\mebi}{Mi}{20}
\journal{NIM-A}

\begin{document}
\begin{frontmatter}

\title{Readout electronics for SUBMET\tnoteref{t1}}
\author[1]{Claudio Campagnari\,\orcidlink{0000-0002-8978-8177}}
\author[2]{Sungwoong Cho\fnref{label1}}
\author[2]{Suyong Choi\,\orcidlink{0000-0001-6225-9876}}
\author[2]{Seokju Chung\,\orcidlink{0009-0004-9134-8527}\fnref{label2}}
\author[3]{Matthew Citron\,\orcidlink{0000-0001-6250-8465}}
\author[4]{Albert De Roeck\,\orcidlink{0000-0002-9228-5271}}
\author[4]{Martin Gastal}
\author[2]{Seungkyu Ha\,\orcidlink{0000-0003-2538-1551}}
\author[5]{Andy Haas}
\author[6]{Christopher Scott Hill\,\orcidlink{0000-0003-0059-0779}}
\author[2]{Insung Hwang\,\orcidlink{0009-0005-9219-8572}\fnref{label3}}
\author[2]{Hoyong Jeong\,\orcidlink{0000-0002-9067-5911}\corref{cor1}}\ead{hoyong5419@korea.ac.kr}
\author[2]{Jaebak Kim\,\orcidlink{0000-0002-2072-6082}}
\author[3]{Jeonghwa Kim}
\author[2]{Hyunki Moon\,\orcidlink{0000-0001-5213-6477}\corref{cor1}}\ead{mhyunki@korea.ac.kr}
\author[1]{Ryan Schmitz\,\orcidlink{0000-0003-2328-677X}}
\author[1]{David Stuart\,\orcidlink{0000-0002-4965-0747}}
\author[2]{Eunil Won\,\orcidlink{0000-0002-4245-7442}}
\author[2]{Jae Hyeok Yoo\,\orcidlink{0000-0003-0463-3043}\corref{cor2}}\ead{jaehyeokyoo@korea.ac.kr}
\author[2]{Jinseok Yoo}
\author[7]{Ayman Youssef}
\author[7]{Ahmad Zaraket}
\author[7]{Haitham Zaraket\,\orcidlink{0000-0003-3929-0431}}

\affiliation[1]{organization={Department of Physics, University of California},
  addressline={Santa Barbara}, 
  city={California},
  citysep={}, 
  postcode={93106}, 
  country={USA}
}
\affiliation[2]{organization={Department of Physics, Korea University},
  addressline={145 Anam-ro, Seongbuk-gu}, 
  city={Seoul},
  citysep={}, 
  postcode={02841}, 
  country={Korea}
}
\affiliation[3]{organization={Department of Physics, University of California},
  addressline={One Shields Avenue, Davis}, 
  city={California},
  citysep={}, 
  postcode={95616}, 
  country={USA}
}
\affiliation[4]{organization={CERN},
  addressline={CH-1211}, 
  city={Geneva},
  country={Switzerland}
}
\affiliation[5]{organization={Department of Physics, New York University},
  addressline={726 Broadway}, 
  city={New York},
  citysep={}, 
  postcode={10012}, 
  country={USA}
}
\affiliation[6]{organization={Department of Physics, The Ohio State University},
  addressline={191 West Woodruff Ave, Columbus}, 
  city={Ohio},
  citysep={}, 
  postcode={43210}, 
  country={USA}
}
\affiliation[7]{organization={Multidisciplinary Physics Lab, Lebanese University},
  addressline={RGHC+4PR}, 
  city={Hadeth-Beirut},
  country={Lebanon}
}

\fntext[label1]{Present address: Department of General Studies, Hongik University, 22639 Sejong-ro, Jochiwon-eup, Sejong, Korea}
\fntext[label2]{Present address: Columbia University, 116th and Broadway, New York, New York 10027, USA}
\fntext[label3]{Present address: Boston University, Commonwealth Ave, Boston, Massachusetts 02215, USA}

\cortext[cor1]{Corresponding authors}
\cortext[cor2]{Principal corresponding author}

\begin{abstract}
A dedicated data acquisition (DAQ) system has been developed for the SUB-Millicharge ExperimenT (SUBMET) at the Japan Proton Accelerator Research Complex (J-PARC), a search for particles carrying a fractional electric charge $Q = \epsilon e$ with $\epsilon$ below $\mathcal{O}(10^{-3})$, hereafter referred to as millicharged particles (mCPs).
Because such particles are expected to produce at most a few scintillation photons, the system is optimized for single-photoelectron detection from the photomultiplier tubes (PMTs), combining high-speed waveform digitization with precise timing.
To capture eight consecutive proton bunches of the \SI{30}{\giga\electronvolt} J-PARC beam within a single trigger, the eight channels of the Domino Ring Sampler 4 (DRS4) chip are cascaded in groups of four to form two readout inputs, each sampling 4096~points continuously at \SI{820.5}{\mega\hertz} over an effective time window of \SI{5}{\micro\second}.
After calibration, timing differences between channels are within \SI{1}{\nano\second} on the same DRS4 chip, \SI{2}{\nano\second} on the same board, and \SI{8}{\nano\second} across different boards, well within the \SI{30}{\nano\second} coincidence window of the experiment.
The front-end electronics achieve an RMS noise below \SI{0.4}{\milli\volt}.
The baseline is deliberately offset upward such that the negative-going pulses span a larger fraction of the digitizer range, improving voltage resolution and dynamic range.
A trigger control board aggregates data from multiple readout boards and sustains the data-transfer rate required for beam operation.
The measured performance confirms that the DAQ system meets the timing, noise, and throughput requirements of the experiment.
\end{abstract}


\begin{highlights}
\item Dedicated DAQ system for SUBMET, a search for millicharged particles.
\item DRS4 cascading to record eight beam bunches per trigger.
\item Continuous sampling of 4096~points at \SI{820.5}{\mega\hertz} over a \SI{5}{\micro\second} window.
\item Timing within \SI{1}{\nano\second} (same chip), \SI{2}{\nano\second} (same board), and \SI{8}{\nano\second} (board-to-board).
\item RMS noise of below \SI{0.4}{\milli\volt} with stable high-rate data transfer.
\end{highlights}


\end{frontmatter}



\section{Introduction}\label{sec:intro}
The search for millicharged particles (mCPs), particles carrying a fractional electric charge $Q = \epsilon e$ with $e$ the elementary charge and $\epsilon \ll 1$, probes physics beyond the Standard Model (SM).
Such particles arise in several SM extensions, for example model that introduces a hidden sector containing a dark photon kinetically mixed with the SM photon.
In the latter case, the hidden-sector states acquire a small effective electric charge and provide viable dark matter candidates~\cite{holdom_u1,edges}.
The SUB-Millicharge ExperimenT (SUBMET) searches for mCPs at the Japan Proton Accelerator Research Complex (J-PARC), where they would be produced in high-intensity proton--target collisions~\cite{submet_proposal}.
Because the energy deposited by a charged particle scales as the square of its charge, an mCP in the sub-millicharge region targeted by SUBMET ($\epsilon \lesssim 10^{-3}$) excites at most a few scintillation photons.
The detector, a scintillator read out by photomultiplier tubes (PMTs), must therefore be sensitive down to single photons.

The technical challenge in SUBMET arises jointly from the J-PARC beam structure and the smallness of the expected signal.
Genuine mCP signals are correlated in time with the proton bunches.
Backgrounds from dark counts and cosmic rays are uncorrelated with the beam and are therefore strongly suppressed by requiring this time correlation, which makes high temporal resolution essential.
Beam-induced backgrounds, in contrast, share the bunch timing and cannot be removed by this requirement; instead, their rate must be predicted precisely and accounted for in the analysis.
High temporal resolution is essential in both cases, both to enforce the correlation with the beam and to resolve the coincidence between layers within a narrow window.

These two requirements are difficult to satisfy simultaneously.
Commercial off-the-shelf DAQ systems generally trade sampling rate against buffer depth, since a higher sampling rate shortens the total recording time that the on-chip buffer and data throughput can sustain.
SUBMET, however, requires both a high sampling rate, to resolve the fast and low-amplitude single-photoelectron pulses, and a long sampling window, to span the full beam spill.
This combination motivated the development of custom readout electronics and a dedicated data acquisition (DAQ) system.
This paper describes the design and implementation of that system, which meets both requirements through an optimized hardware architecture and high-speed data management.

The remainder of this paper is organized as follows.
Section~\ref{sec:submet} gives an overview of SUBMET, the J-PARC beam characteristics relevant to the experiment, and the multi-channel DAQ strategy.
Section~\ref{sec:require} discusses the constraints imposed by the J-PARC beam and the detector that drive the system specifications.
Sections~\ref{sec:hw} and~\ref{sec:fw} detail the hardware and firmware architecture, and Section~\ref{sec:perf} presents the performance validation results.

\section{SUBMET at J-PARC}\label{sec:submet}
The feasibility of searching for mCPs in \SI{30}{\giga\electronvolt} proton fixed-target collisions at J-PARC has been studied previously~\cite{submet_proposal}.
In that study, mCPs are produced in the decays of neutral mesons, namely $\pi^0$, $\eta$, $\rho$, $\omega$, $\phi$, and $J/\psi$, created in the proton--target collisions.

The J-PARC beam used for neutrino production~\cite{t2k} is delivered in a characteristic time structure of eight bunches per spill~\cite{jparc_mr}, which strike the target sequentially~\cite{jparc_beam}.
Each bunch is about \SI{34}{\nano\second} wide with an inter-bunch spacing of \SI{581}{\nano\second}, and the eight bunches therefore extend over roughly \SI{4}{\micro\second}.
The mCPs produced in neutral-meson decays inherit this eight-bunch structure and travel along the beam direction.
The Neutrino Monitor building, used for neutrino-beam monitoring, is located about \SI{280}{\meter} downstream of the target.
The SUBMET detector sits on its second basement level and searches for mCPs arriving in these eight bunches.

\begin{figure}[h]
\begin{center}
\includegraphics[width=1.0\linewidth]{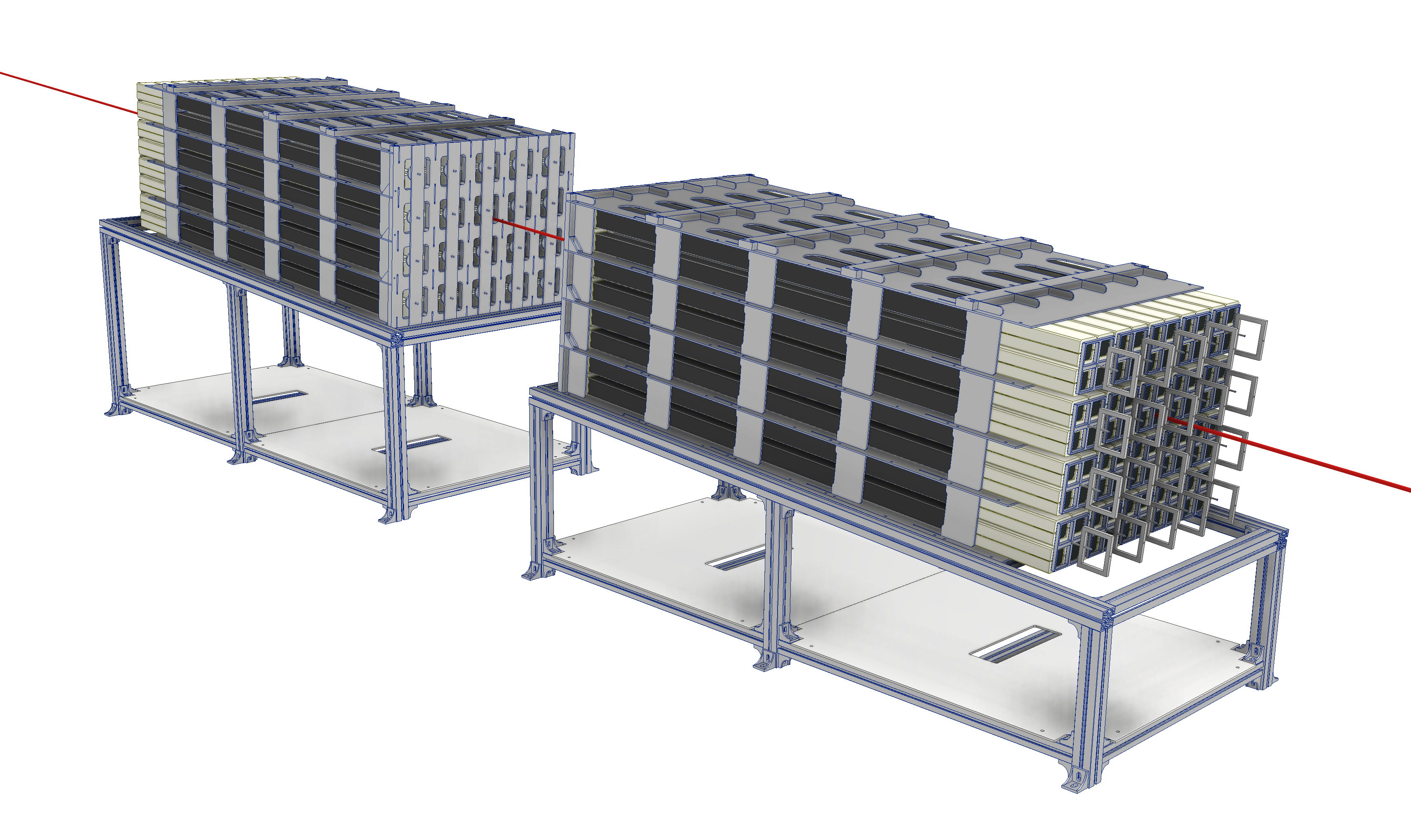}
\end{center}
\caption{A 3D model of the SUBMET detector.
Each module consists of a \qtyproduct{50 x 50 x 1500}{\milli\meter} plastic scintillator bar coupled to a PMT.
The detector comprises two such layers, aligned such that the track of an mCP (red line) passes through both layers within a narrow time window.}
\label{fig:detector}
\end{figure}

To be sensitive to mCPs, which interact only feebly with matter, the SUBMET detector is built from plastic scintillator bars elongated along the mCP momentum direction.
To detect the photons of scintillation by mCPs, a photomultiplier tube (PMT) is coupled to each bar.
The combination of one scintillator bar and one PMT is referred to as a module.
Each bar measures \qtyproduct{50 x 50 x 1500}{\milli\meter}, and a module therefore covers a transverse area of \qtyproduct{50 x 50}{\milli\meter}.
Eighty modules are arranged in a \numproduct{10 x 8} stack to cover the transverse spread of mCPs from the target, and a second, identical stack is placed behind the first along the beam direction, giving a total of 160 modules in two layers.
Figure~\ref{fig:detector} shows the resulting detector structure.

For every beam collision with the target, a beam-synchronous trigger initiates the readout, in which the PMT signals of all 160 modules are digitized and recorded without any signal-based requirement, that is, as a zero-bias readout.

The scintillation light produced by an mCP scales with the square of its charge.
For the charge of interest by the experiment, $\epsilon < 10^{-3}$, the scintillation-light yield from an mCP is sufficiently low that the number of photoelectrons detected by the PMT is of order one, $\mathcal{O}(1)$.
In this regime, either zero or one photoelectron is recorded in most cases, even when an mCP traverses the detector. The readout is therefore designed to detect a single-photoelectron (SPE) pulse as the expected signal.
Larger charges yield proportionally more photoelectrons and hence multiple-SPE pulses, which are correspondingly easier to detect; the SPE case therefore represents the most demanding requirement on the readout.
With all PMT bias voltages set to \SI{1.3}{\kilo\volt}, an SPE pulse from the Hamamatsu R7725 PMTs used here has a width of \qtyrange{5}{20}{\nano\second} in full width at half maximum and an amplitude of \qtyrange{5}{30}{\milli\volt}, measured as the peak height relative to the pedestal, as determined by their gain and output characteristics.

An mCP candidate is identified as a spatial and temporal coincidence of SPE-like pulses across the two layers.
Two collinear modules, one in each layer and aligned along the beam direction, must each register an SPE-like pulse.
In the offline analysis, the expected time-of-flight between the layers, about \SI{8}{\nano\second}, is first subtracted from the second-layer hit time, and the two pulses are then required to coincide within a residual window.
This window does not represent the time-of-flight itself but accommodates the overall timing spread of the detected pulses.
The dominant contribution is the variation in the optical path length of scintillation photons inside the \SI{1500}{\milli\meter} bar, since photons emitted at different points and in different directions propagate and reflect over different distances before reaching the PMT.
For the Eljen EJ-200 scintillator used here ($n = 1.58$)~\cite{submet_mech}, an axial round trip of about \SI{3}{\meter} corresponds to a propagation time of roughly \SI{16}{\nano\second}, and oblique paths can be longer.
In practice, photons travelling much beyond the \SI{3.8}{\meter} attenuation length of EJ-200 are attenuated before reaching the PMT.
Adding the single-photoelectron transit-time spread of the PMT and the electronics timing offsets, the coincidence window is set conservatively to \SI{30}{\nano\second} in order to retain genuine coincidences.

\section{Experimental Requirements and Specifications}\label{sec:require}
Spatial alignment of the detector is a mechanical matter~\cite{submet_mech} and is not addressed here.
The timing precision, by contrast, is determined by the readout and data acquisition (DAQ) system.
This section translates the J-PARC beam structure, the detector geometry, and the PMT signal characteristics into quantitative requirements on the DAQ system, and explains why these requirements call for custom-developed electronics rather than a commercial digitizer.

As described in Section~\ref{sec:submet}, the J-PARC neutrino beam is delivered in spills of eight proton bunches, each about \SI{34}{\nano\second} wide and separated by \SI{581}{\nano\second}, which makes a full spill span roughly \SI{4}{\micro\second}.
To monitor the time-of-flight and the correlation of each hit with the beam timing, all eight bunches must be captured within a single acquisition cycle.
The readout system must therefore provide a continuous sampling window of at least \SI{5}{\micro\second} per trigger, leaving margin around the \SI{4}{\micro\second} bunch train.

The signal characteristics impose the complementary requirements on sampling speed and resolution.
The detector comprises two layers of 80 modules each, giving 160 PMT channels that must be digitized simultaneously for every trigger.
The expected signal is a SPE pulse from a Hamamatsu R7725 PMT.
At the \SI{1.3}{\kilo\volt} operating bias, this pulse is about \qtyrange{5}{20}{\nano\second} wide in full width at half maximum, with an amplitude of \qtyrange{5}{30}{\milli\volt} measured as the peak height relative to the pedestal.
These properties set three requirements.
First, to reconstruct the pulse shape, at least four samples must fall on the narrowest (\SI{5}{\nano\second}) pulse, which requires a sampling interval of \SI{1.25}{\nano\second} or shorter.
Second, the amplitude resolution must resolve the \qtyrange{5}{30}{\milli\volt} pulses, together with the sub-millivolt features needed to separate an SPE pulse from the baseline, across a dynamic range of about \SI{1}{\volt}.
This requires a digitization resolution of at least 12~bit, corresponding to \SI{0.25}{\milli\volt} per least significant bit.
Third, the pedestal noise must remain below \SI{1}{\milli\volt} RMS in order to keep SPE pulses clearly distinguishable from baseline fluctuations with high detection efficiency.

Table~\ref{tab:requirements} summarizes the requirements derived above.
Their combination is demanding, since a high sampling rate, needed for the fast and low-amplitude pulses, and a long record length, needed for the microsecond-scale bunch train, must be achieved simultaneously across 160 channels.
\begin{table}[h]
\centering
\caption{Requirements on the SUBMET DAQ system and the parameters that drive them.}
\label{tab:requirements}
\begin{tabular}{lll}
\hline
Parameter            & Requirement                            & Driving consideration                      \\
\hline \hline
Sampling window      & $\geq \SI{5}{\micro\second}$           & 8 bunches over $\sim\SI{4}{\micro\second}$ \\
Sampling interval    & $\leq \SI{1.25}{\nano\second}$         & $\geq 4$ samples per SPE pulse             \\
Buffer depth         & $\geq 4000$ cells                      & window / interval                          \\
Amplitude resolution & $\geq 12$~bit over $\sim\SI{1}{\volt}$ & resolve \qtyrange{5}{30}{\milli\volt} SPE  \\
Pedestal noise       & $< \SI{1}{\milli\volt}$ RMS            & separate SPE from baseline                 \\
Channel count        & 160, simultaneous                      & 2 layers $\times$ 80 modules               \\
\hline
\end{tabular}
\end{table}

These requirements guided the choice of the sampling device.
The Domino Ring Sampler 4 (DRS4)\footnote{\url{https://www.psi.ch/en/drs}} was selected accordingly.
The DRS4 is a switched-capacitor array providing nine channels, each \num{1024} cells deep, with a sampling rate that can be set between about \SI{0.7}{\giga\hertz} and \SI{5}{\giga\hertz}.
A single \num{1024}-cell channel, even at the lowest rate, spans only about \SI{1.5}{\micro\second}, far short of the \SI{5}{\micro\second} window.
The chip is therefore operated in a cascaded configuration, in which four channels are chained into one \num{4096}-cell input, yielding two such inputs per chip.
Distributing \num{4096} cells across the \SI{5}{\micro\second} window fixes the sampling rate at approximately \SI{820.5}{\mega\hertz}, corresponding to a \SI{1.22}{\nano\second} interval.
This rate lies within the DRS4 operating range, exceeds the \num{4000}-cell depth requirement, and still places about four samples on the narrowest \SI{5}{\nano\second} pulse.
The sampled waveforms are subsequently digitized by a 12-bit ADC, as detailed in Section~\ref{sec:hw}.

At the system level, the 160 channels and the resulting data volume are managed by dividing the readout into ten readout boards (ROBs) and a single trigger control board (TCB).
Each ROB digitizes 16 channels, performing the high-speed sampling and local buffering.
The TCB acts as the master controller, synchronizing the ROBs and aggregating their data for transmission to the central server.
Per trigger, each channel produces \num{4096} samples.
Although the ADC digitizes each sample to 12~bit, the samples are stored as 16-bit words for technical reasons, and one channel therefore occupies \qty{8}{\kibi\byte} per trigger, including a two-sample event header.
One ROB consequently generates approximately \qty{128}{\kibi\byte} per trigger.
Under nominal beam operation the ten ROBs together deliver on the order of \qty{1}{\mebi\byte\per\second} to the TCB, which the TCB must receive and forward to the server.
Because the aggregation is fully pipelined in the firmware, this rate lies well within the capability of the TCB rather than posing a limiting constraint.
The implementation is described in Section~\ref{sec:fw:summary}.

The combination of these requirements distinguishes SUBMET from other mCP searches, which have adopted readout systems matched to their own beam environments.
The milliQan experiment at the Large Hadron Collider (LHC)~\cite{milliqan} and the proposed FORMOSA experiment~\cite{formosa} employ commercial high-speed digitizers such as the CAEN V1743\footnote{\url{https://caen.it/products/v1743/}} to capture fast scintillator pulses.
Such digitizers offer multi-GS/s sampling but a limited record length per channel, which suits the \SI{25}{\nano\second} bunch spacing of the LHC.
The J-PARC spill, by contrast, distributes its eight bunches over several microseconds, and the same devices would therefore not cover the full bunch train in a single acquisition.
The SUBMET system is instead optimized to satisfy all three constraints together, namely a high sampling rate, a microsecond-scale window, and high channel density, in a compact and cost-effective form.
This balance motivated the custom development described in the following sections.

\section{Design of the Readout Electronics}\label{sec:hw}
As introduced in Section~\ref{sec:require}, the SUBMET readout electronics form a hierarchical, distributed system that combines high-speed sampling with large-scale data aggregation.
It comprises two functional units, the readout board (ROB) and the trigger control board (TCB).
\begin{itemize}
\item \textbf{Readout board (ROB).} Ten ROBs are deployed, each sampling and digitizing 16 PMT channels with Domino Ring Sampler 4 (DRS4) chips and performing local data processing and buffering.
\item \textbf{Trigger control board (TCB).} A single TCB distributes a common clock and synchronized triggers to the ten ROBs and aggregates their digitized data, buffering it in a Double Data Rate 4 (DDR4) memory before the formatted packets are sent to the back-end server.
\end{itemize}
The two-layer detector geometry drives the channel mapping.
As shown in Figure~\ref{fig:detector}, an mCP candidate appears as a coincidence between two collinear modules, one in each layer.
To measure the relative timing of such a pair as precisely as possible, the two modules of each collinear pair are cabled to the two inputs of a single DRS4 chip, such that their signals are sampled on the same switched-capacitor array against a common timing reference.
As shown in Section~\ref{sec:perf}, this yields sub-nanosecond relative timing within a pair, well matched to the coincidence requirement.

The ten ROBs and the TCB are connected in a star topology by optical fibers.
Over these links the TCB broadcasts the system clock and trigger to keep all 160 channels sampling in phase, while each ROB streams its digitized waveforms back to the TCB.
This modular arrangement provides scalable channel density and precise timing synchronization.
The following subsections describe the two units in detail, beginning with the analog signal path on the ROB.

\subsection{Readout Board (ROB)}\label{sec:hw:rob}
The analog front-end of the ROB is designed to preserve the fast timing and small amplitude of the detector signals.
Each PMT output enters the ROB through a SubMiniature version A (SMA) connector terminated in \SI{50}{\ohm} for impedance matching.
The single-ended PMT signal is then converted to a differential signal by a THS4520 (Texas Instruments) fully differential amplifier (FDA), a high-speed and ultra-low-noise device selected to keep the noise floor below \SI{0.4}{\milli\volt} RMS.
The PMT signal is applied to the non-inverting input while the inverting input is tied to analog ground.
This differential conversion suppresses common-mode noise and electromagnetic interference, and it matches the differential input of the DRS4 that follows.
The differential pair is routed over length-matched, impedance-controlled traces to minimize timing skew before reaching the digitizer.

The differential signals are sampled by the DRS4, a switched-capacitor analog ring buffer.
In SUBMET the DRS4 samples at \SI{820.5}{\mega\hertz}, a rate set by the FPGA firmware, as described in Section~\ref{sec:fw}.
Each DRS4 channel comprises \num{1024} storage cells, and at \SI{820.5}{\mega\hertz} a single channel therefore covers only \SI{1.249}{\micro\second}, far too short for the eight-bunch spill that spans about \SI{4}{\micro\second}.
Four channels are therefore cascaded into one continuous \num{4096}-cell sampler, extending the window to approximately \SI{5}{\micro\second}.
Commercial DRS4-based digitizers do not provide this cascading mode, which is why a custom ROB was developed.

The DRS4 has nine channels, comprising eight signal channels and one auxiliary channel.
With four-channel cascading, the eight signal channels form two \num{4096}-cell inputs, and, as noted above, these two inputs are wired to the two modules of a collinear coincidence pair.
The ninth channel does not carry detector signals.
Instead, it receives a timing-calibration pattern generated by the FPGA, which corrects the DRS4 cell timing when it drifts (Section~\ref{sec:fw}).
Each ROB thus carries eight DRS4 chips to serve its 16 analog inputs, that is, eight collinear pairs.

On a trigger, the domino wave is stopped and the voltages held in the DRS4 cells are read out sequentially for digitization.
This readout is far slower than the sampling.
The frozen cells are digitized by a 12-bit ADC (AD9222, Analog Devices) at \SI{10.4167}{\mega\hertz}, obtained by dividing the \SI{125}{\mega\hertz} reference clock by twelve.
In other words, \SI{820.5}{\mega\hertz} is the analog sampling (write) rate into the DRS4, whereas \SI{10.4167}{\mega\hertz} is the rate at which the stored samples are read out and converted.
The AD9222 provides eight channels per chip.
To digitize the 16 inputs of a board with one 8-channel ADC, each DRS4 multiplexes its two cascaded paths onto a single output, and the eight streams from the eight DRS4 chips are converted by the eight ADC channels.

The readout blocks further sampling and therefore determines the dead time of the acquisition.
During the write state the DRS4 samples continuously, so the dead time is set solely by the digitization of the stored cells.
Each ADC channel reads out the two cascaded inputs of a chip in sequence, that is, \num{8192} cells at \SI{10.4167}{\mega\hertz}:
\begin{equation}
t_\mathrm{dead} \approx \frac{\num{8192}}{\SI{10.4167}{\mega\hertz}} \approx \SI{786}{\micro\second}.
\end{equation}
The corresponding maximum trigger rate of about \SI{1.3}{\kilo\hertz} lies far above the rates required by the experiment, and the dead time therefore does not limit the acquisition, as confirmed in Section~\ref{sec:perf}.

Control and data processing on the ROB are handled by a Xilinx Kintex-7 FPGA (XC7K70T-1FBG484C) together with a CoolRunner-II CPLD (XC2C128-7VQG100C).
The FPGA is the primary controller.
It manages the DRS4 timing, generating the domino-wave sampling clock, stopping the sampling on a trigger, and sequencing the readout, and it supplies the clock and control signals for the AD9222.
The eight serialized ADC streams are routed to the FPGA, where they are aggregated for transmission to the TCB.
The FPGA also drives the auxiliary ninth DRS4 channel with a dedicated timing-calibration pattern through a DAC.
The structure of this pattern and its role in timing calibration are described in Section~\ref{sec:fw}.

The CPLD runs from its own \SI{50}{\mega\hertz} oscillator, independent of the FPGA clock.
Its main task is to set the DRS4 analog offset.
Through a digital-to-analog converter (DAC, TLV5638) and an operational amplifier (MCP6001), it generates the DC offset applied during readout.
This offset realizes the aforementioned elevated baseline.
Because the PMT pulses are negative-going, with the negatively biased PMT sourcing electrons as the photoelectron cascade is amplified, raising the baseline allows each pulse to swing across a larger fraction of the ADC input range, which improves the effective voltage resolution and dynamic range for these small signals.
The CPLD can also accept an independent external trigger through an SMA connector and buffer.
Although the board normally uses the TCB-distributed trigger, this input provides redundancy for standalone testing and special trigger conditions.

\begin{figure*}[h]
\centering
\includegraphics[width=0.99\textwidth]{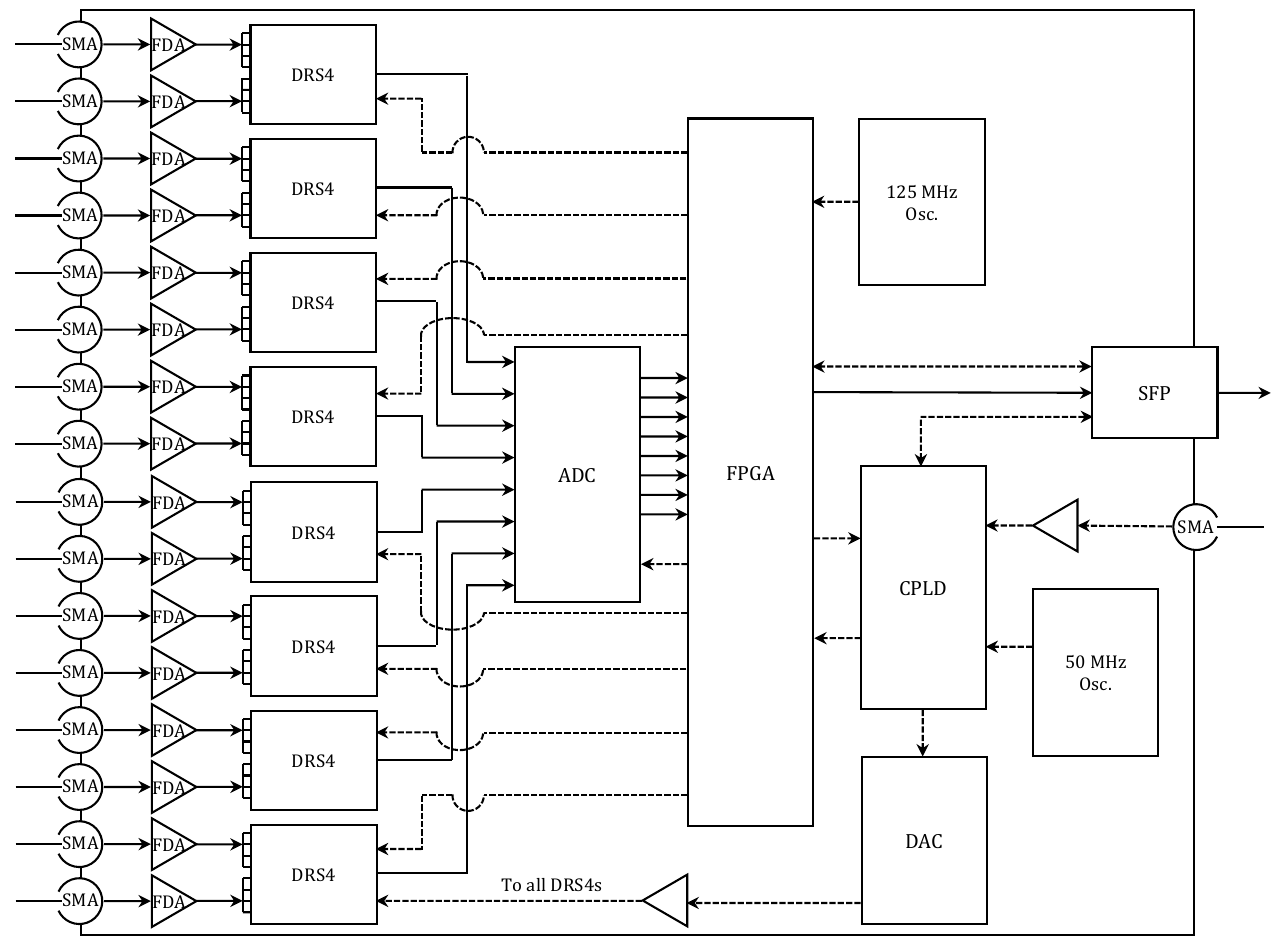}
\caption{\label{fig:rob}Block diagram of the ROB hardware architecture.
The diagram illustrates the signal path from the PMT inputs through the analog front-end to the DRS4 and ADC stages.
Solid lines denote the primary data flow, including analog signals and digitized data, while dashed lines represent control, timing, and synchronization signals managed by the FPGA and CPLD.}
\end{figure*}

For board-to-board communication and synchronization, each ROB uses a small form-factor pluggable (SFP) optical module as the physical interface to the optical fiber.
The module is driven by an on-chip GTX gigabit transceiver of the FPGA\footnote{\url{https://docs.amd.com/r/en-US/ug440-xilinx-power-estimator/Using-the-Transceiver-Sheets-GTP-GTX-GTH-GTY-GTZ}}, which serializes the outgoing data and recovers the clock and data from the incoming stream.
Over this optical link the ROB performs hardware identification, clock synchronization, and reception of the TCB-distributed triggers, keeping all boards phase-aligned across the DAQ system.
The resulting ROB architecture and the interconnection of these components are shown in the block diagram of Figure~\ref{fig:rob}.

\subsection{Trigger Control Board (TCB)}\label{sec:hw:tcb}
The TCB is the central controller of the DAQ system, coordinating trigger distribution and aggregating data from the ROBs.
It is built around a Xilinx Artix-7 FPGA (XC7A200T-2FFG1156C) that manages high-speed serial communication, data buffering, and the external trigger interface.
The TCB connects to the ten ROBs over optical fibers through SFP modules.
Its front panel carries 13 SFP ports, all wired to the FPGA multi-gigabit transceivers, of which ten serve SUBMET and three are reserved for future expansion.
Incoming waveforms are checked for integrity, namely packet length and header, before being written to the onboard \qty{4}{\gibi\byte} DDR4 synchronous dynamic random-access memory (SDRAM), which acts as a circular buffer that absorbs the high-rate bursts of a beam spill.

Timing and configuration are handled by a dedicated clocking sub-section.
A \SI{125}{\mega\hertz} crystal oscillator and a CDCE62005RGZ (Texas Instruments) clock generator provide the base clocks for the FPGA and a CPLD (XC95144XL-10TQ144C).
The CPLD manages the boot sequence, initialization, and FPGA configuration.

A key function of the TCB is synchronization with the J-PARC Main Ring (MR).
The TCB receives an external trigger through a LEMO connector and the corresponding beam-spill number through an emitter-coupled logic (ECL) connector, both from the MR control system.
The spill number is a 17-bit integer that increments with every spill.
These signals pass through low-voltage differential signaling (LVDS) receivers to internal logic levels before reaching the FPGA, which timestamps each trigger and tags it with its spill number for precise event reconstruction.

\begin{figure*}[h]
\centering
\includegraphics[width=0.99\textwidth]{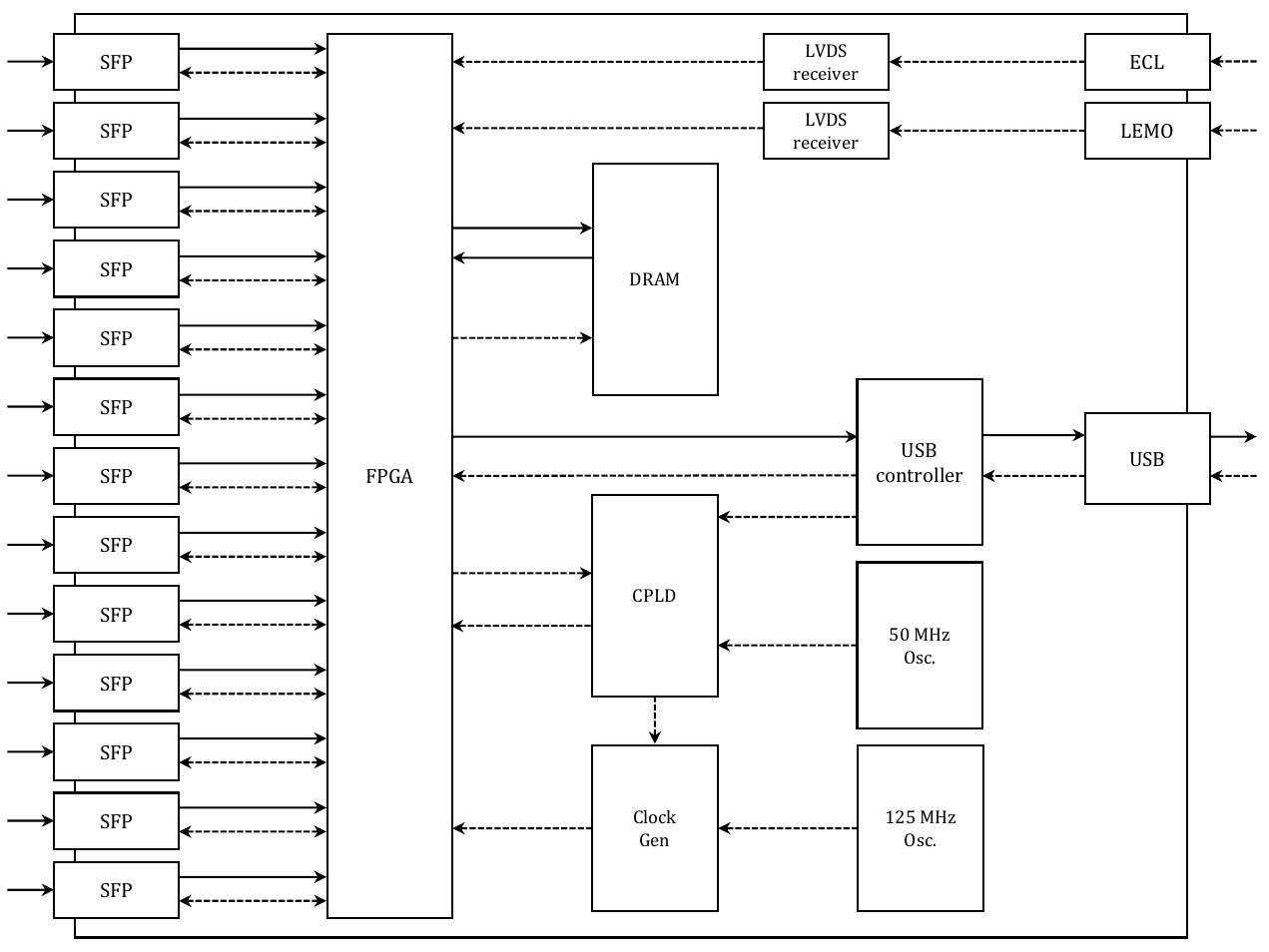}
\caption{\label{fig:tcb}Block diagram of the TCB hardware architecture. The central FPGA manages the data aggregation from the ROBs via 13 SFP ports and handles the external trigger/spill signals from the J-PARC MR. The \qty{4}{\gibi\byte} DDR4 SDRAM acts as a high-speed buffer for event data before transmission to the server via the USB interface.}
\end{figure*}

Data transfer to the back-end server is handled by a universal serial bus (USB) controller interfaced to the FPGA and CPLD.
The FPGA reads the buffered data from the DDR4 memory and streams them to the server in a first-in--first-out manner, and the same interface lets the server issue configuration and monitoring commands to the TCB.
Figure~\ref{fig:tcb} summarizes the internal structure of the TCB and its interfaces for trigger input, board-to-board communication, and data output.

\section{Firmware Design and Data Format}\label{sec:fw}
\subsection{Timing Synchronization and Clock Distribution}\label{sec:fw:sync}
The identification of coincidence events across spatially distributed detector modules depends critically on the precision of the timing synchronization.
The DAQ system meets this requirement through a hierarchical clock-distribution network in which the TCB serves as the central controller.
The TCB encodes a \SI{125}{\mega\hertz} reference clock together with control data into a \SI{2.5}{\giga\bit\per\second} serial stream and broadcasts it to each ROB over an optical fiber through the GTX transceivers.
The serialization applies 8b/10b encoding, which preserves DC balance and guarantees a sufficient transition density for reliable clock-and-data recovery (CDR) in the receiving transceiver.
Each ROB carries its own \SI{125}{\mega\hertz} oscillator as a local reference, and the phase-locked loop (PLL) within the FPGA recovers the \SI{125}{\mega\hertz} clock embedded in the optical stream.
The recovered clock is then synchronized in frequency and phase to this local reference, establishing a common time base between the TCB and every ROB.

Because each ROB may begin its clock counter at a different instant during power-up, a static timing offset can arise between boards.
This offset is bounded by the \SI{8}{\nano\second} period of the \SI{125}{\mega\hertz} clock and, once the link is locked, remains constant throughout operation.
An initial calibration measures these offsets after power-up, and they are subsequently removed in the offline analysis to establish a unified time base across the detector.
The residual timing spread achieved after this calibration is presented in Section~\ref{sec:perf}.

The FPGA derives the domino-wave clock that operates the DRS4 sampling ring from the same \SI{125}{\mega\hertz} base.
One domino-wave period spans 312 cycles of the \SI{125}{\mega\hertz} clock, giving a domino-wave frequency of approximately \SI{400.6}{\kilo\hertz}.
Each period advances the domino wave through two of the \num{1024} storage cells, so a full \num{1024}-cell channel is sampled in \num{512} periods, or about \SI{1.25}{\micro\second}.
The resulting effective sampling rate is $2 \times \num{1024} \times \SI{400.6}{\kilo\hertz} \approx \SI{820.5}{\mega\hertz}$\footnote{The exact value is \SI{820.5128}{\mega\hertz}, given by $2 \times \num{1024} \times \SI{125}{\mega\hertz} / 312$.}.

\subsection{Trigger Logic and Sequential Data Flow}\label{sec:fw:trg}
\begin{sidewaysfigure*}[p]
\centering
\includegraphics[width=0.99\textwidth]{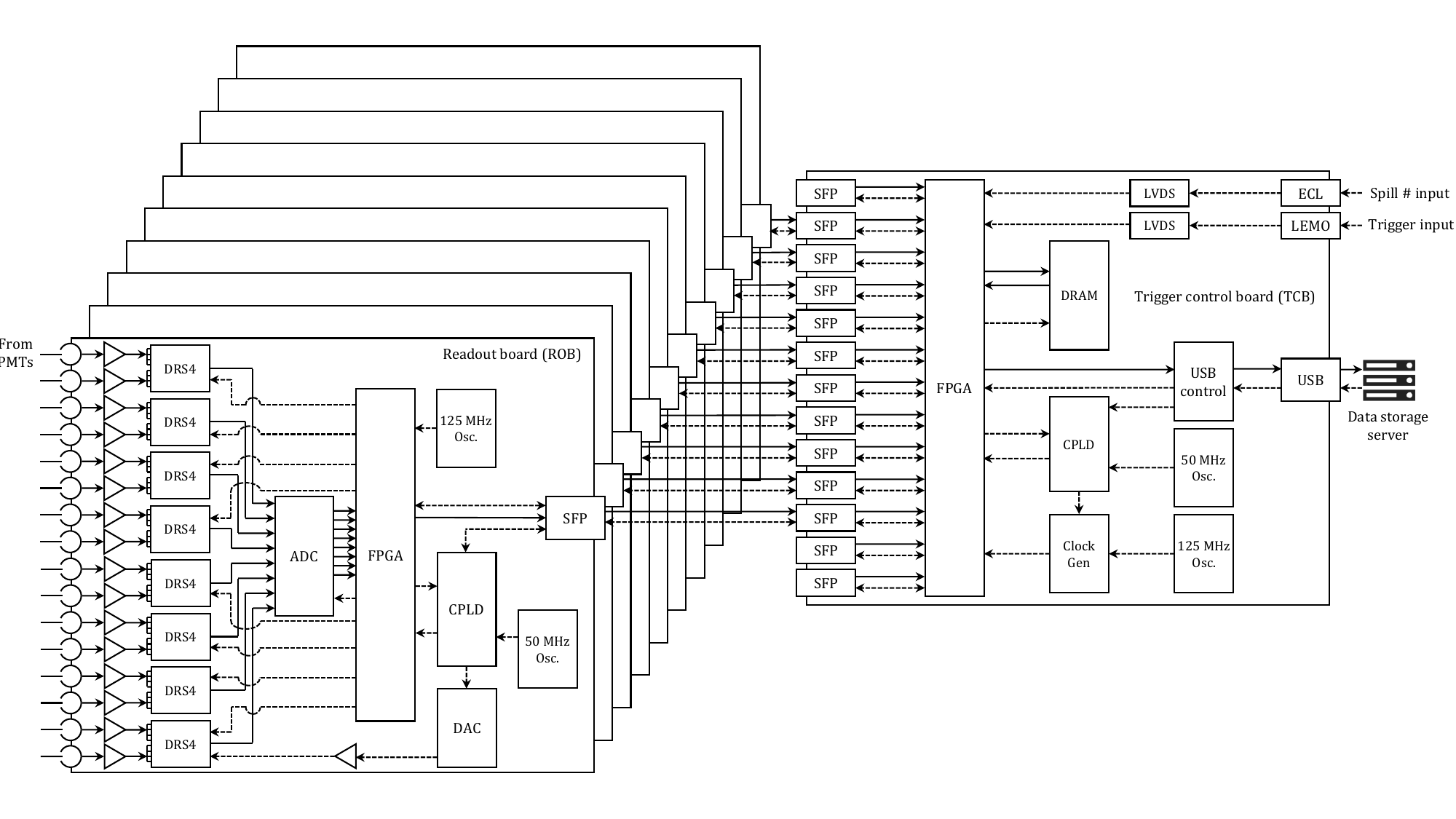}
\caption{\label{fig:sig_proc} Illustration of the signal processing in the ROBs and dataflow to the TCB.}
\end{sidewaysfigure*}

The acquisition cycle is governed by the TCB, which handles both software-generated triggers issued through the USB interface and hardware-timed triggers received from the J-PARC Main Ring.
When a valid trigger condition is satisfied, the TCB broadcasts a trigger packet to all ROBs.
Upon receiving the trigger, each ROB terminates the DRS4 sampling process in order to preserve the charge stored in the switched-capacitor arrays.
The FPGA thereupon initiates a sequential readout in which the stored voltages are digitized by the AD9222 ADC.
All components within a ROB operate from a common FPGA clock, which constrains the spread in trigger-stop timing among the DRS4 chips to the nanosecond level.
The corresponding measurement is reported in Section~\ref{sec:perf}.
After digitization, the data are encapsulated in a dedicated packet format and transmitted to the TCB.
The TCB verifies the integrity of each incoming packet and stores it in the DDR4 buffer, which is operated as a first-in--first-out (FIFO) memory able to hold up to two complete event windows.
This buffering absorbs the high-rate data bursts produced during a beam spill before the events are transferred to the DAQ computer over USB.
The signal processing in the principal board components is illustrated in Figure~\ref{fig:sig_proc}.

\subsection{Detailed Data Packet Format}\label{sec:fw:data}
The packet structure is designed for straightforward alignment verification and for future scalability.
Each event from a single ROB produces a fixed-size packet of \qty{128}{\kibi\byte} (\num{131072}~bytes).
The data-length field is consequently constant at 0x20000, which simplifies buffer management in the back-end server.
Table~\ref{tab:datapacket} specifies the header fields and the waveform payload.
\begin{table*}
\begin{center}
\begin{tabularx}{\textwidth}{l l X}
 \hline
 Contents           &          Size & Properties                                                           \\
 \hline\hline
 Data length        &       4 bytes & the length of data                                                   \\
 ROB ID             &       1 byte  & ROB index                                                            \\
 Spill number       &       3 bytes & 17-bit beam spill number from the MR, stored in a 3-byte field       \\
 TCB trigger time   &       8 bytes & trigger time counted in TCB                                          \\
 TCB trigger number &       4 bytes & trigger count at TCB                                                 \\
 DRS4 stop time     &       8 bytes & DRS4 stop timing at ROBs                                             \\
 ROB trigger number &       4 bytes & trigger count at ROB                                                 \\ 
 DRS4 stop address  &      16 bytes & Stop position of capacitor in DRS4s                                  \\
 DRS4 PLL stat      &       1 byte  & Check bits whether DRS4 phase loop locked                            \\
 Reserved           &       7 bytes & Reserved bits for future upgrade                                     \\
 trigger time, sec  &       4 bytes & Time of arrival measured in the computer in second                   \\
 trigger time, msec &       4 bytes & Time of arrival measured in the computer in millisecond              \\
 Data               & 131,008 bytes & 4,094 samples $\times$ 2~bytes $\times$ 16 channels of waveform data \\
 \hline
\end{tabularx}
\caption{\label{tab:datapacket} Specification of the ROB-to-TCB data packet format.
The first two data samples from 16 channels are replaced to the header.
All multi-byte fields are stored in little-endian byte order, and the waveform samples are organized as 16-bit words.}
\end{center}
\end{table*}
Several fields secure the robustness of the data stream.
The ROB ID and the ROB trigger number enable the software to detect and correct misalignments that may arise when a packet from a particular board is delayed or lost.
In addition to the precise TCB trigger time, the packet carries trigger-time fields in seconds and milliseconds that are filled by the DAQ computer when the data arrive over USB, which allows the internal experiment clock to be correlated with the wall-clock time.

Each 12-bit ADC sample is stored in a 16-bit word to align the waveform data with the byte-oriented memory of the back-end system.
The first two samples of every channel are overwritten by the event header, and the retained \num{4094} samples per channel therefore yield a waveform payload of \num{131008}~bytes (\numproduct{4094 x 2 x 16}).

\subsection{System Throughput and Data Volume Summary}\label{sec:fw:summary}
The TCB constitutes the central node of the DAQ hierarchy.
It distributes the synchronized triggers to the ROBs, aggregates the individual data streams, and manages the final transfer to the DAQ computer.
Each cascaded DRS4 input acquires \num{4096} samples per trigger.
The first two samples are replaced by header information, and the remaining \num{4094} samples constitute the waveform payload.
Because the substituted header occupies the same space, the data recorded to disk retain the full \qty{8192}{bytes} per input, and the 16 inputs of a ROB produce the \qty{128}{\kibi\byte} packet described above.
Across the full 160-channel detector, a single event amounts to approximately \qty{1.25}{\mebi\byte}.
For the J-PARC T2K beam cycle of \SI{1.28}{\second}, one event per spill corresponds to a steady-state rate of about \qty{1.0}{\mebi\byte\per\second}.
These figures reflect the nominal beam cycle rather than any limit of the system.
The throughput actually sustained at much shorter trigger intervals is characterized in Section~\ref{sec:perf}.

\section{Performance Evaluation}\label{sec:perf}
\subsection{Pedestal and Noise Performance}\label{sec:perf:noise}
The sensitivity of SUBMET to mCPs depends strongly on the signal-to-noise ratio of the readout electronics.
The baseline noise performance was first estimated analytically by summing the contributions of the principal active components, namely the FDA (THS4520), the sampler (DRS4), and the ADC (AD9222).
The THS4520 exhibits an input voltage noise of \SI{2}{\nano\volt\per\sqrt\hertz}, which remains well below the millivolt level once integrated over the system bandwidth.
The DRS4, operating as a switched-capacitor array, contributes an intrinsic noise of approximately \SI{0.35}{\milli\volt} RMS.
Together with the quantization noise of the AD9222, the theoretical lower limit of the electronic noise for the complete readout chain was estimated to be approximately \SI{0.4}{\milli\volt} RMS or below.

The noise performance was subsequently measured under realistic experimental conditions.
The PMT outputs were connected to the ROB through SMA cables, and the full bias high voltage was applied to the PMTs in order to include any noise pickup from the detector setup.
The pedestal level was recorded across all 160 channels over multiple beam spills.
\begin{figure}[t]
\centering
\includegraphics[width=0.7\textwidth]{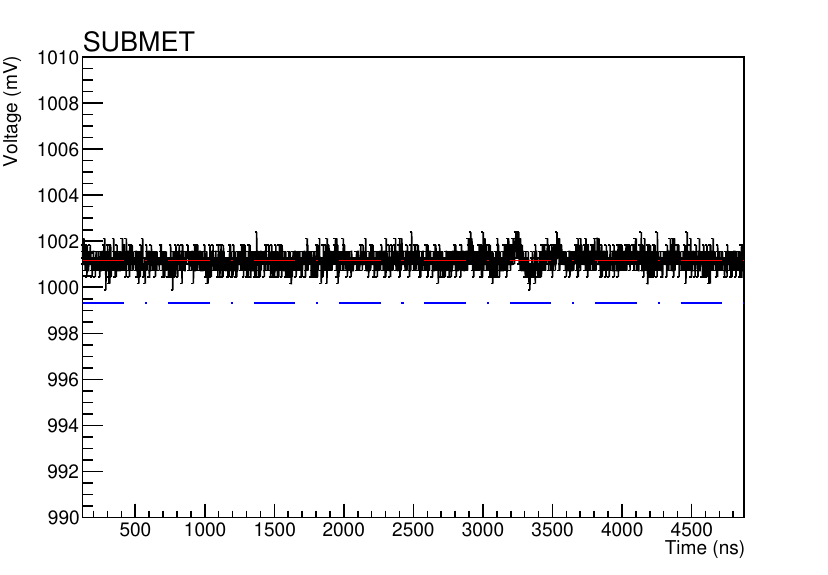}
\caption{\label{fig:pedestal}A typical pedestal distribution from one channel.
The offset voltage is set to approximately \SI{1}{\volt} to secure the dynamic range for the negative-going pulses.
The RMS of this channel is \SI{0.37}{\milli\volt}.
The red dotted line marks the pedestal mean, \SI{1001.17}{\milli\volt}, and the blue dotted line marks the pulse-detection threshold, set five times the RMS below the mean.}
\end{figure}
Figure~\ref{fig:pedestal} presents a typical pedestal distribution for a single channel.
The baseline follows a Gaussian profile, and the measured noise floor of this channel is \SI{0.37}{\milli\volt} RMS, below the \SI{0.4}{\milli\volt} target.
To confirm that this performance is uniform across the detector, the pedestal RMS was measured for all 160 channels over \num{10000} events, and the mean RMS of each channel was computed.
\begin{figure}[t]
\centering
\includegraphics[width=0.7\textwidth]{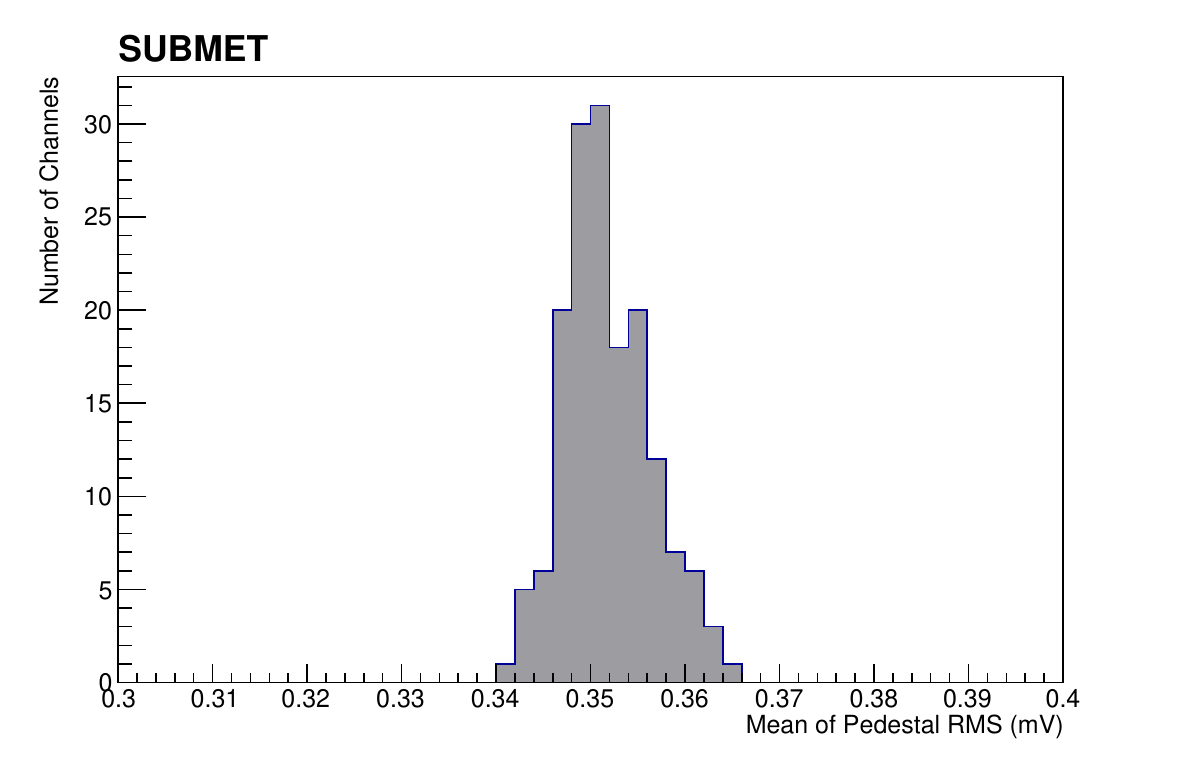}
\caption{\label{fig:pedrms}Distribution of the mean pedestal RMS of the 160 readout channels.
For each channel, the pedestal RMS was evaluated over \num{10000} events and averaged.
All channels lie within \SIrange{0.34}{0.37}{\milli\volt}, below the \SI{0.4}{\milli\volt} target and confirming a uniform, low noise floor across the detector.}
\end{figure}
Figure~\ref{fig:pedrms} shows the distribution of these per-channel mean values, which lie within \SIrange{0.34}{0.37}{\milli\volt} across the full detector.
For a given channel, the RMS fluctuates by at most about $\pm\SI{0.02}{\milli\volt}$ around its own mean over the \num{10000} events.
The noise therefore remains below the \SI{0.4}{\milli\volt} target in every channel and for every event.
This uniformly low noise agrees with the analytical estimate and confirms the effectiveness of the low-noise design, in particular the fully differential signal path.
During operation, the pulse-detection threshold is computed for each event as five times the pedestal RMS of the corresponding channel, set below the pedestal mean in order to select the negative-going SPE pulses.
On average this threshold corresponds to approximately \SI{1.8}{\milli\volt} below the baseline.
Defining the threshold relative to the measured noise keeps the false-trigger probability from Gaussian pedestal fluctuations negligible.
With this setting, the measured SPE detection efficiency is 0.995 for the Hamamatsu R7725 PMTs, the fraction of the SPE pulse-height distribution retained above the threshold~\cite{lp2023}.
The pedestal was monitored over extended periods and exhibited negligible drift, ensuring stable thresholds during long-term operation.

\subsection{Timing Synchronization}\label{sec:perf:sync}
The principal physics objective of SUBMET, the identification of coincidence events between the two layers, requires nanosecond-level timing precision.
As described in Section~\ref{sec:fw}, the design targeted an in-chip timing difference below \SI{1}{\nano\second}, a chip-to-chip difference within \SI{2}{\nano\second} on the same board, and a board-to-board difference within \SI{8}{\nano\second}.
The in-chip term is the most important of the three.
As introduced in Section~\ref{sec:hw}, the two modules of each collinear pair are cabled to a pair of channels on the same DRS4 sampler, which preserves the relative timing of coincidence candidates at the hardware level.
\begin{figure*}[h]
\centering
\includegraphics[width=0.99\textwidth]{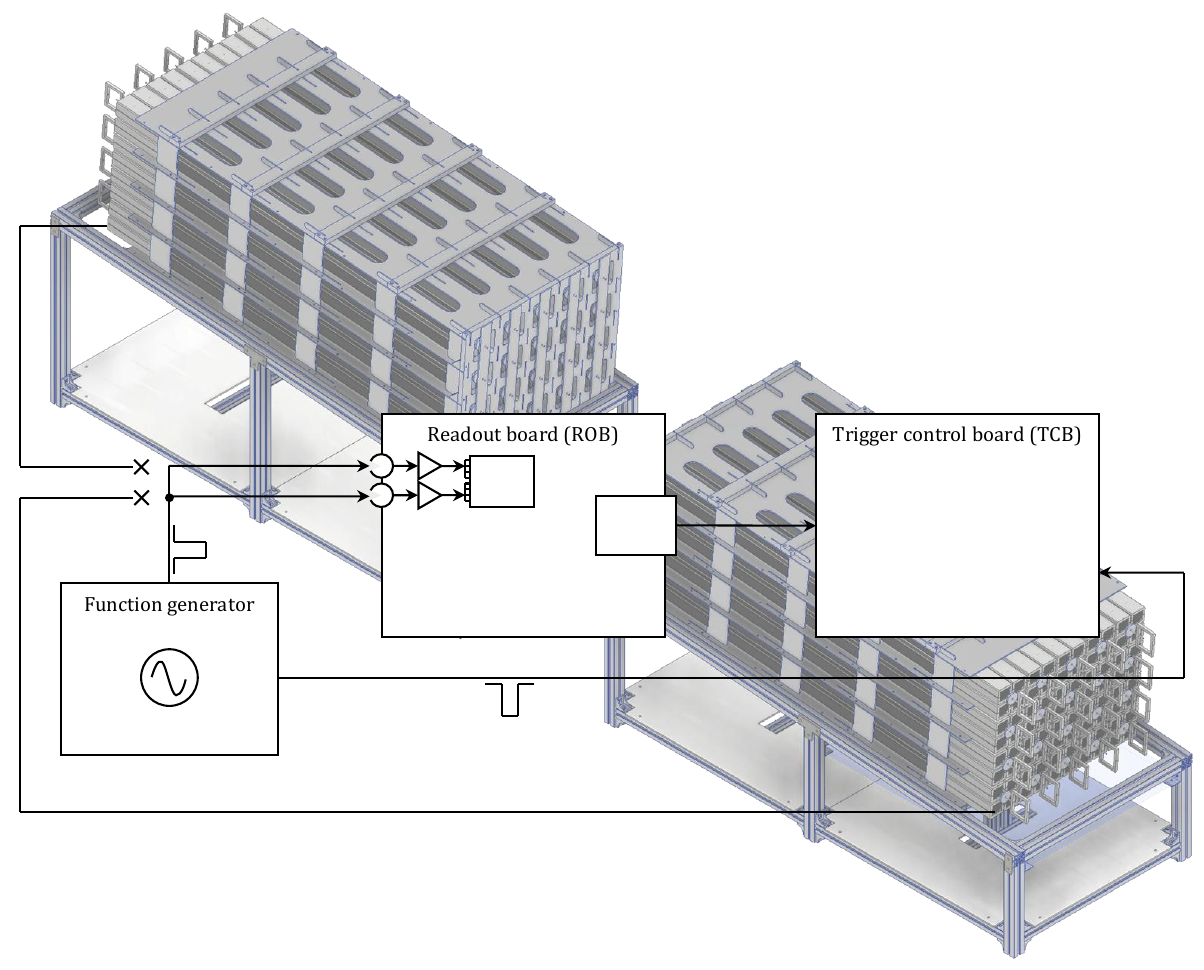}
\caption{\label{fig:timingcal}A test setup for measuring the time difference between adjacent channels for coincidence event measurements of a module pair. Square waves generated by a function generator are simultaneously injected into both channels, and a synchronous trigger is input to the TCB. The detector structure is depicted in Ref.~\cite{submet_mech}.}
\end{figure*}
These specifications were verified in a bench test using the setup shown in Figure~\ref{fig:timingcal}.
The readout system contains 80 channel pairs that share a DRS4 sampler.
A function generator injected identical rectangular pulses, with a \SI{10}{\nano\second} rise and fall time, a \SI{100}{\nano\second} width, and a \SI{400}{\milli\volt} amplitude, into the two channels under test.
A trigger synchronized with these pulses was simultaneously supplied to the TCB to initiate acquisition.
For each of the 80 pairs, the timing difference $\Delta t$ was analyzed over \num{10000} trigger events.
\begin{figure}[h]
\centering
\includegraphics[width=0.8\textwidth]{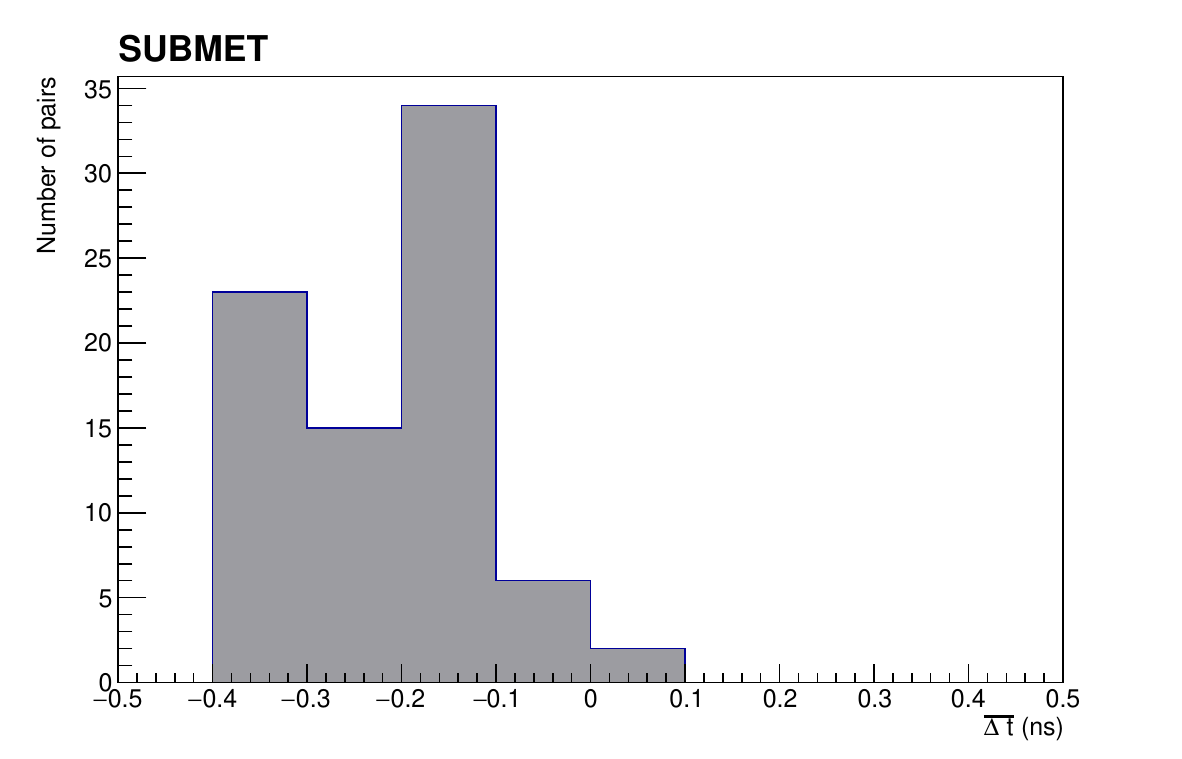}
\caption{\label{fig:dtdist}Distribution of $\overline{\Delta t}$ measurements between 80 module pairs.}
\end{figure}
Figure~\ref{fig:dtdist} shows the distribution of the mean timing difference $\overline{\Delta t}$ for the 80 pairs.

The mean values cluster near zero, between about \SI{-0.4}{\nano\second} and \SI{0.1}{\nano\second}, with a slight bias toward negative values.
This offset is attributed to differences in the PCB trace lengths between the two channels of a pair, introduced at the artwork stage along the paths from the analog input to the sampler, from the sampler to the ADC, and from the ADC to the FPGA.
Being static and reproducible, this shift is removed by the timing calibration and does not affect coincidence identification.
Relative to the sampling period of \SI{1.22}{\nano\second} (\SI{820.5}{\mega\hertz}), these sub-nanosecond deviations are negligible and confirm the in-chip synchronization.
The chip-to-chip and board-to-board differences were measured to be within \SI{2}{\nano\second} and \SI{8}{\nano\second}, respectively, in agreement with the design values.
All three terms are well within the \SI{30}{\nano\second} coincidence window defined in Section~\ref{sec:submet}, which confirms that the timing performance of the readout system is sufficient for mCP identification.

\subsection{Data Transfer Stability and Throughput Analysis}\label{sec:perf:data_rate}
The full readout system operates with ten ROBs in parallel.
Each event produces a fixed payload of \num{131072}~bytes per ROB, giving an aggregate event size of
\begin{equation}
10 \times \num{131072}~\text{bytes} = \num{1310720}~\text{bytes} = \qty{1.25}{\mebi\byte}.
\end{equation}
Under nominal physics operation, one event is recorded per spill.
For the J-PARC fast-extraction cycle of \SI{1.28}{\second}, this corresponds to an event rate of about \SI{0.8}{\hertz} and a data rate of approximately \qty{1.0}{\mebi\byte\per\second}, amounting to roughly \qty{82}{\gibi\byte} per day.
When a control sample is additionally recorded between spills in dedicated runs, both the rate and the daily volume double, reaching approximately \qty{1.95}{\mebi\byte\per\second} and \qty{165}{\gibi\byte} per day.

The system sustains rates well beyond these physics requirements.
The ROB-to-TCB links use GTX optical transceivers at multi-gigabit line rates, which far exceed the data rate of the experiment and therefore do not constrain the throughput.
The relevant limit arises instead at the server storage, whose sequential write speed bounds the sustained acquisition rate; in the present configuration this is a RAID array of 7200~RPM hard disk drives.
Long-term tests confirmed continuous, loss-free acquisition at event rates up to \SI{50}{\hertz}, corresponding to
\begin{equation}
50 \times \qty{1.25}{\mebi\byte} = \qty{62.5}{\mebi\byte\per\second}.
\end{equation}
This demonstrated capability exceeds the nominal physics rate of \SI{0.8}{\hertz} by nearly two orders of magnitude, providing a substantial margin for the needs of SUBMET. The rate could be raised further with faster storage, such as NVMe solid-state drives.

\section{Summary}\label{sec:summary}
We have developed a scalable, high-performance data acquisition system for SUBMET, the search for millicharged particles at J-PARC.
The architecture comprises ten synchronized readout boards (ROBs) and a central trigger control board (TCB), and it meets the stringent requirements of low-noise waveform sampling and precise timing synchronization.
The performance of the system was validated in a series of bench tests.
Using the DRS4 switched-capacitor array together with low-noise fully differential amplifiers, the readout chain achieves an electronics noise floor below \SI{0.4}{\milli\volt} RMS, uniform across all 160 channels.
The pulse-detection threshold is set for each channel at five times the pedestal RMS below the baseline, corresponding to about \SI{2}{\milli\volt}.
With this noise performance, a single-photoelectron detection efficiency of 0.995 was measured for the Hamamatsu R7725 PMTs.
The system digitizes \num{4096} samples per channel at a sampling rate of \SI{820.5}{\mega\hertz} across all 160 channels simultaneously, providing an effective time window of \SI{5}{\micro\second} that captures the full eight-bunch spill and enables reconstruction of the faint PMT signals.
Timing precision, which is essential for identifying genuine two-layer coincidences and suppressing uncorrelated backgrounds, was measured at the nanosecond level.
The relative timing between two channels sharing a DRS4 sampler was found to be below \SI{1}{\nano\second}, while the chip-to-chip and board-to-board differences remained within \SI{2}{\nano\second} and \SI{8}{\nano\second}, respectively.
Because the two modules of each collinear pair are read out by the same sampler, this sub-nanosecond in-chip precision applies directly to coincidence candidates, and all three terms lie well within the \SI{30}{\nano\second} coincidence window of the experiment.
Under nominal operation the system records one event per spill, corresponding to about \SI{0.8}{\hertz} for the \SI{1.28}{\second} J-PARC fast-extraction cycle.
Long-term tests demonstrated stable, loss-free acquisition at a sustained rate of \SI{50}{\hertz}, far above this physics rate.
The throughput is currently limited by the sequential write speed of the server storage rather than by the FPGA logic or the USB 3.0 interface, both of which retain substantial headroom for higher-rate operation.
The modular design also provides room for expansion.
Of the 13 optical ports on the TCB, ten serve the present detector and three remain available, permitting growth to 13 ROBs (208 channels) without hardware modification.
The successful implementation and verification of this readout system represent an important milestone for SUBMET.
The demonstrated low-noise and high-precision timing performance provides the sensitivity required to explore a previously inaccessible region of millicharged-particle parameter space, reaching charges down to $\epsilon \sim 10^{-3}$ for masses up to approximately \SI{1.6}{\giga\electronvolt\per\clight\squared}.

\section*{Acknowledgements}
The authors are grateful to Sangyeol Kim for his substantial support in the design and fabrication of the data acquisition system. We also thank the T2K Collaboration for providing the beam trigger and spill signal information, which was incorporated into the system design.

We acknowledge the following funding agencies that support the investigators who carried out this research in various capacities: Swiss Funding Agencies (Switzerland); DOE and NSF (USA); Lebanese University (Lebanon). This work was partially supported by the National Research Foundation of Korea (NRF) grants funded by the Korea government (MSIT) (RS-2021-NR059935 and RS-2025-00560964) and a Korea University grant.
This research was also supported by Basic Science Research Program through the NRF funded by the Ministry of Education (RS-2026-25559812).



\bibliographystyle{elsarticle-num-names}
\bibliography{refs.bib}

\end{document}